\documentclass[12pt]{article}

\newlength{\vshift}
\newlength{\hshift}
\usepackage{amssymb,amsopn}

\begin{document}

\begin{titlepage}
\begin{flushright}
ICMPA-MPA/2009/13\\
\end{flushright}

\begin{center}
{\ }\vspace{1cm}

{\Large\bf { Twisted Grosse-Wulkenhaar $\phi^{\star 4}$ model: dynamical noncommutativity and Noether currents}}
%\vspace{5pt}
 %{\Large\bf of Generalized Models with Spin-Orbit Interaction}

 %\vspace{1.5cm}
\end{center}

\begin{center}
Mahouton Norbert Hounkonnou$^{\dag }$ and Dine Ousmane Samary$^{\textasteriskcentered}$
\end{center}
\vspace{1cm}
\begin{center}
{\em $^\dag$ International Chair of Mathematical Physics
and Applications} \\
{\em ICMPA-UNESCO Chair}\\
{\em University of Abomey-Calavi}\\
{\em 072 B.P. 50 Cotonou, Republic of Benin}\\
{\em $^{\textasteriskcentered}$ Institut de Math\'ematiques et de Sciences Physiques (IMSP)}\\
{\em University of Abomey-Calavi}\\
{\em 01 B.P. 613 Porto-Novo, Republic of Benin}\\
{\em E-mail: {\tt $^{\dag}$ norbert.hounkonnou@cipma.uac.bj\footnote{Correspondence author (with copy to hounkonnou@yahoo.fr).}

{\tt $^{\textasteriskcentered}$ ousmanesamarydine@yahoo.fr }}}
\end{center}
\vspace{1.0cm}

\today

%\clearpage
\begin{abstract}
This paper addresses the computation of Noether currrents
for the renormalizable  Grosse-Wulkenhaar (GW) $\phi^{\star 4}$ model
subjected to a dynamical
noncomutativity realized through a twisted Moyal product. The noncommutative (NC) energy-momentum tensor (EMT),
 angular momentum tensor (AMT) and the dilatation current (DC) are explicitly derived.
   The breaking of translation and rotation invariances has been avoided  via a constraint equation.
\end{abstract}

%\pacs{02.40.Gh, 11.10.Nx.%03.65.-w, 03.65.Ca, 03.65.Ge, 02.30.Sa}
%}
% \submitto{\JPA}
% \today
 %\maketitle

%%%%%%%%%%%%%%%%%%%%%%%%%%%%

%%%%%%%%%%%%%%%%%%%%%%%%%%%%
\section{Introduction}
Most of different settings for noncommutative (NC) field theories \cite{aschieri} - \cite{Langmann} are based
  on a Moyal space
${\rm I\!\!R}_\Theta^D$, a deformed D-dimensional space endowed with a constant Moyal $\star-$bracket of
coordinate functions
 \begin{eqnarray}
\left[x^\mu, x^\nu\right]_{\star} = i \Theta^{\mu\nu}
\end{eqnarray}
where $\Theta$ is a $D\times D$ non-degenerate skew-symmetric matrix (which requires D even),
 usually  chosen in the form
\begin{eqnarray}
\Theta = \left(\begin{array}{llllllll}
0 & \Theta_{1} &  &  &  &  &  &  \\
-\Theta_{1}  &0  &  &  &  &0  &  &  \\
 &  & 0 &\Theta_{2}  &  &  &  &  \\
 &  &- \Theta_{2} & 0 &  &  &  &  \\
 &  &  &  &\vdots & \vdots &  &  \\
 &  &  &  & \vdots & \vdots &  &  \\
 &  &0  &  &  &  & 0 &\Theta_{\frac{D}{2}}  \\
 &  &  &  &  &  & -\Theta_{\frac{D}{2}} & 0
\end{array}\right)
\end{eqnarray}
where $\Theta_{j}\in {\mathbb{R}}$, $j=1,2,\cdots, \frac{D}{2}$,
have dimension\footnote{Units such that $\hbar=1=c$
are used throughout.} of length square, ($[\Theta_j] = [L]^2$),
$D$ denoting the spacetime dimension.
 The corresponding product of functions is the associative, noncommutative
Moyal-Groenewold-Weyl product,
simply called hereafter {Moyal product or $\star$-product}
defined by
\begin{eqnarray}\label{stproduct}
(f\star g)(x) = {\rm m}\left\{
e^{i \frac{\Theta^{\rho\sigma}}{2}
 \partial_{\rho}\otimes \partial_{\sigma}} f(x)\otimes g(x)
 \right\}, \,
 x\in {\rm I\!\!R}_\Theta^D, \qquad  \forall f,g\in\mathcal{S}({\rm I\!\!R}_\Theta^D)
 \end{eqnarray}
${\rm m}$ is the   ordinary multiplication of functions and
 $\mathcal{S}({\rm I\!\!R}_\Theta^D)$ - the space of suitable Schwartzian functions.
For more details, see [11]-[14].
  Such a noncommutative
 geometry possesses the  specific pathology to break both the Lorentz invariance by the presence of  $\Theta^{\mu\nu}$,
 as $\left[x^\mu, x^\nu\right]_{\star} = i \Theta^{\mu\nu}$ is not generally invariant under rotation, and the
 local character of the theory due to infinite time derivatives. There result  energy momentum tensors (EMTs)
  which are not locally conserved,
not traceless in the massless situation and,
not symmetric and not gauge invariant in gauge theories.
A number of works exist in attempts to achieve regularization for the NC EMT which then becomes symmetric
albeit not locally conserved. Further improvement of this quantity by usual algebraic  tricks breaks its symmetry,
(see \cite{Ben-Houk1}  and references therein).  Therefore, the property of  nonlocal conservation of angular momentum
is not a priori proscribed.

Recently, Paolo Aschieri {\it et al} \cite{aschieri} introduced a so-called dynamical noncommutativity to investigate Noether currents in an ordinary nonrenormalizable twisted $\phi^{\star 4}$ theory. This work addresses  questions of the applicability of such a formalism on  the new class of
renormalizable NC field theories (NCRFT) built on the Grosse and Wulkenhaar (GW)
$\phi^{\star 4}$ scalar field model defined in Euclidean space-time  by the action functional \cite{Grosse}
%{\tiny
\begin{eqnarray} \label{GW}
 S_{\star}^{\Omega}[\phi]&=&\int \mbox{d}^{D}x \,\Big(\frac{1}{2}
\partial_{\mu}\phi\star\partial^{\mu}\phi+
\frac{\Omega^{2}}{2}(\tilde{x}_{\mu}\phi)
\star(\tilde{x}^{\mu}\phi)+\frac{m^{2}}{2}\phi
\star\phi
\cr 
&&+\frac{\lambda}{4!}\phi\star\phi\star\phi\star\phi\Big),
\end{eqnarray}
%}
where $\tilde{x}_{\mu}=2(\Theta^{-1})_{\mu\nu} x^{\nu}$ and $S_{\star}^{\Omega}[\phi]$
 is covariant under Langmann-Szabo duality \cite{Langmann}.
  $\Omega$ and $\lambda$ are  dimensionless parameters.
Generalizing the Moyal $\star$-product (\ref{stproduct})  under the form
\begin{eqnarray}
(f\star g)(x)=m\Big\{e^{i\frac{\Theta^{ab}}{2}X_{a}\otimes X_{b}}f(x)\otimes g(x)\Big\}=: e^\Delta (f,g)
\end{eqnarray}
where $X_{a}=e_{a}^{\mu}(x)\partial_{\mu}$ is a commuting vector
fields, the commutation relation becomes
$[x^{\mu},x^{\nu}]_{\star}=i\Theta^{ab}e_{a}^{\mu}(x)e_{b}^{\nu}(x)
=:i\widetilde{\Theta}^{\mu\nu}(x)$, engendering a twisted scalar
field theory where $e_{a}^{\mu}$,
 and hence the $\star$ product itself, appear dynamical. See Appendix for useful relations concerning
 this generalized product.
  The condition $[X_{a},X_{b}]=0$
 implies constraints on $e_{a}^{\mu}$, namely $e_{[a}^{\nu}\partial_{\nu}e_{b]}^{\mu}=0$,
  that can be solved off-shell in terms of $D$ scalar fields $\phi^{a}$,
 (see \cite{aschieri} and \cite{paolo}).
   Supposing that the square matrix $e_{a}^{\mu}$ has an inverse $e_{\mu}^{a}$ everywhere, so that
   the  $X_{a}$ are linearly independent,
    then the above condition becomes $\partial_{[\mu}e_{\nu]}^{a}=0$ which is satisfied by $e_{\nu}^{a}=\partial_{\nu}\phi^{a}$.
    Besides, the Leibniz rule extends to the commuting fields $X_{a}$ as follows:
       $X_{a}(f\star g)=(X_{a}f)\star g+f\star(X_{a}g)$.

This paper is organized as follows. In Section 2, we derive the field equations of motion and
 provide with the explicit computation of  noncommutative energy momentum tensor (NC EMT),
  angular momentum tensor (AMT) and dilatation current (DC). Furtheremore, we proceed to
  the symmetry analysis including the translation,  rotation and
 dilatation transformations and compute the conserved currents. Finally, we end with some concluding
 remarks in Section 3.

\section{Twisted Grosse-Wulkenhaar model: Noether currents}
The generalized NC GW Lagrangian action corresponding to (\ref{GW}) can be written as:
\begin{eqnarray}\label{1}
 \mathcal{S}_{\star}^{\Omega}[\phi]&=&\int\,ed^{D}x\,\,\Big(\mathcal{L}_{\star}^{\Omega}\star e^{-1}\Big)\cr
&=&\int\, e{\mbox{d}}^D x\hspace{1mm} \Big\{\frac{1}{2}\partial_{\mu}\phi\star\partial^{\mu}\phi
+\frac{m^2}{2}\phi\star\phi+\frac{\lambda}{4!}\phi\star\phi\star\phi\star\phi\cr &&+\frac{\Omega^{2}}{2}(\tilde{x}_{\mu}\phi)\star(\tilde{x}^{\mu}\phi)
+\frac{1}{2}\partial_{\mu}\phi_{a}\star\partial^{\mu}\phi^{a}\Big\}\star e^{-1},
%\\
%&\equiv &\mathcal{S}_{\star}^{v,\phi}+\mathcal{S}_{\star}^{ m}
%+\mathcal{S}_{\star}^{\lambda}+\mathcal{S}_{\star}^{\Omega,\phi}
% +\mathcal{S}_{\star}^{v,\phi^{a}}\nonumber
\end{eqnarray}
$$\mbox{ where} \quad e=\mbox{det}e^{a}_{\mu}$$
from which the peculiar Euler Lagrange equations of motion can be
readily derived by direct application of the variational principle. There results the following.
\begin{enumerate}
\item[i)]  From the
$\phi$-variation, the equation of motion of the field $\phi$ is expressed as:
\begin{eqnarray}\label{em1}
\mathcal{E}_{\phi}&=&-\frac{1}{2}\partial_{\sigma}\Big(e\{\partial^{\sigma}\phi,e^{-1}\}_{\star}\Big)
+\frac{m^{2}}{2}e\{\phi,e^{-1}\}_{\star}+\frac{\lambda}{4!}e\{\phi\star\phi,\{\phi,e^{-1}\}_{\star}\}_{\star}\cr
&+&\frac{\Omega^{2}}{8}
e\{\tilde{x},\{e^{-1},\{\tilde{x},\phi\}_{\star}\}_{\star}\}_{\star}=0.
\end{eqnarray}
In the commutative limit $\Theta \rightarrow 0$, the equation (\ref{em1}) becomes the usual $\phi^4$ field equation of
motion
\begin{eqnarray}
 \square \phi - m^2 \phi - {\lambda \over {3!}}\phi^3=0.
\end{eqnarray}
The current $\mathcal{K}^{\sigma}$ is determined by the expression
\begin{eqnarray}
 \mathcal{K}^{\sigma}=\mathcal{K}^{\sigma}(0)+\mathcal{K}^{\sigma}(m^2)
+\mathcal{K}^{\sigma}(\lambda)+\mathcal{K}^{\sigma}(\Omega^2),
\end{eqnarray}
where the four main contributions are induced by
the velocity field
\begin{eqnarray}
 \mathcal{K}^{\sigma}(0)&=& \frac{e\delta\phi}{2}.\{\partial^{\sigma}\phi,e^{-1}\}_{\star}+ee_{b}^{\sigma}\Big[T(\Delta)\Big(\delta\partial_{\mu}\phi,\frac{\widetilde{X}^{b}}{2}\{\partial^{\mu}\phi,e^{-1}\}_{\star}\Big)\nonumber\\
&&+S(\Delta)\Big(\partial_{\mu}\phi,\widetilde{X}^{b}(\partial^{\mu}\delta\phi\star e^{-1})\Big) \Big],
\end{eqnarray}
the mass term
\begin{eqnarray}
\mathcal{K}^{\sigma}(m^2)&=&ee_{b}^{\sigma}\Big[\frac{m^2}{2}T(\Delta)
\Big(\delta\phi,\widetilde{X}^{b}\{\phi,e^{-1}\}_{\star}
\Big)\cr 
&&+m^{2}S(\Delta)\Big(\phi,\widetilde{X}^{b}(\delta\phi\star
e^{-1})\Big)\Big],
\end{eqnarray}
the $\phi^{\star 4}$ interaction
\begin{eqnarray}
\mathcal{K}^{\sigma}(\lambda)&=& ee_{b}^{\sigma}\Big[\frac{\lambda}{4!}T(\Delta)\Big(\delta\phi,\widetilde{X}^{b}\{\phi\star\phi,\{\phi,e^{-1}\}_{\star}\}_{\star}\Big)\cr
&&+\frac{\lambda}{12}S(\Delta)\Big(\phi,\widetilde{X}^{b}(\delta\phi\star\phi\star\phi\star e^{-1})\Big)
\cr
&& +\frac{\lambda}{12}S(\Delta)\Big(\phi\star\phi,\widetilde{X}^{b}(\delta\phi\star\phi\star e^{-1})\Big)\cr &&+\frac{\lambda}{12}S(\Delta)\Big(\phi\star\phi\star\phi,\widetilde{X}^{b}(\delta\phi\star e^{-1})\Big)\Big]
\end{eqnarray}
and the GW harmonic interaction
\begin{eqnarray}
\mathcal{K}^{\sigma}(\Omega^2)&=& ee_{b}^{\sigma}\Big[\frac{\Omega^{2}}{8}T(\Delta)\Big(\delta\phi,\widetilde{X}^{b}\{\tilde{x},\{e^{-1},\{\tilde{x},\phi\}_{\star}\}_{\star}\}_{\star}\Big)\nonumber\\
&&+\frac{\Omega^{2}}{4}S(\Delta)\Big(\tilde{x},\widetilde{X}^{b}(\delta\phi\star\{\tilde{x},\phi\}_{\star}\star e^{-1})\Big)\cr
&&+\frac{\Omega^{2}}{4}S(\Delta)\Big(\{\tilde{x},\phi\star\tilde{x}\}_{\star},X^{b}(\delta\phi\star e^{-1})\Big)\nonumber\\
&&+\frac{\Omega^{2}}{4}S(\Delta)\Big(\{\phi,\tilde{x}\}_{\star},\widetilde{X}^{b}
(\delta\phi\star\tilde{x}\star e^{-1})\Big)\Big],
\end{eqnarray}
 respectively. See Appendix for definitions and notation.
\item[ii)] From the {$\phi^{c}$-variation}, after tedious algebraic transformations, we get the following field 
equation:
\begin{eqnarray}
\mathcal{E}_{(\phi,\phi^{c})}&=&e\Big[\frac{1}{e}X_{c}(\mathcal{L}_{\star}^{\Omega})- (X_c\phi)\Big(\frac{m^2}{2}\{\phi, e^{-1}\}_{\star}+\frac{\lambda}{4!}\{\phi\star\phi,\{\phi,e^{-1}\}_{\star}\}_{\star}\cr
&&+\frac{\Omega^{2}}{2}\tilde{x}.\{\tilde{x}\phi,e^{-1}\}_{\star}\Big)-\frac{\Omega^{2}}{2}\phi X_{c}\tilde{x}.\{\tilde{x}\phi,e^{-1}\}_{\star}\cr &&-\frac{1}{2}X_{c}\partial_{\mu}\phi.\{\partial^{\mu}\phi,e^{-1}\}_{\star}-\frac{1}{2}X_{c}\partial_{\mu}\phi_{a}.\{\partial^{\mu}\phi^{a},e^{-1}\}_{\star}\cr
&&-\frac{1}{e}\partial_{\mu}\Big(\frac{e}{2}\{\partial^{\mu}\phi_{c},e^{-1}\}_{\star}
\Big]=0.\nonumber\\
\end{eqnarray}
Using  the identities $
\tilde{x}_{\mu}\star\phi=\tilde{x}_{\mu}\phi+i
\partial_{\mu}\phi\quad\mbox{ and }\quad\phi\star\tilde{x}_
{\mu}=\tilde{x}_{\mu}\phi-i\partial_{\mu}\phi$
implying $\tilde{x}\phi=\frac{1}{2}\{\tilde{x},\phi\}_{\star}$, we
can deduce that  $\frac{\Omega^{2}}{2}\tilde{x}.
\{\tilde{x}\phi,e^{-1}\}_{\star}=\frac{\Omega^{2}}{8}\{\tilde{x},
\{e^{-1},\{\tilde{x},\phi\}_{\star}\}_{\star}\}_{\star}$, and
the equation of motion takes the form
\begin{eqnarray}
 \mathcal{E}_{(\phi,\phi^{c})}&=&-X_{c}\phi\mathcal{E}_{\phi}+X_{c}\mathcal{L}_{\star}^{\Omega}-\frac{1}{2}X_{c}\phi\partial_{\mu}\Big(e\{\partial^{\mu}\phi,e^{-1}\}_{\star}\Big)\cr
&&-e\frac{\Omega^{2}}{2}\phi X_{c}\tilde{x}.\{\tilde{x}\phi,e^{-1}\}_{\star}-\frac{e}{2}X_{c}\partial_{\mu}\phi.\{\partial^{\mu}\phi,e^{-1}\}_{\star}\cr &&-\frac{e}{2}X_{c}\partial_{\mu}\phi_{a}.\{\partial^{\mu}\phi^{a},e^{-1}\}_{\star}-\partial_{\mu}\Big(\frac{e}{2}\{\partial^{\mu}\phi_{c},e^{-1}\}_{\star}\Big)\nonumber\\
&=& -X_{c}\phi\mathcal{E}_{\phi}-\mathcal{E}_{\phi^{c}}=0,
\end{eqnarray}
where
\begin{eqnarray}\label{em2}
\mathcal{E}_{\phi^{c}}&=& -X_{c}\mathcal{L}_{\star}^{\Omega}+\frac{1}{2}X_{c}\phi\partial_{\mu}\Big(e\{\partial^{\mu}\phi,e^{-1}\}_{\star}\Big)+e\frac{\Omega^{2}}{2}\phi X_{c}\tilde{x}.\{\tilde{x}\phi,e^{-1}\}_{\star}\nonumber\\
&&+\frac{e}{2}X_{c}\partial_{\mu}\phi.\{\partial^{\mu}\phi,e^{-1}\}_{\star}+\frac{e}{2}X_{c}\partial_{\mu}\phi_{a}.\{\partial^{\mu}\phi^{a},e^{-1}\}_{\star}\cr
&&+\partial_{\mu}\Big(\frac{e}{2}\{\partial^{\mu}\phi_{c},e^{-1}\}_{\star}\Big)
\end{eqnarray}
with
$$\frac{\Omega^{2}}{2}\phi X_{c}\tilde{x}.\{\tilde{x}\phi,e^{-1}\}_{\star}=\frac{\Omega^{2}}{8}X_{c}\tilde{x}.
\{\phi,\{e^{-1},\{\tilde{x},\phi\}_{\star}\}_{\star}\}_{\star}.$$
One can immediately show that, as expected from \cite{aschieri}, when $\phi$ is on shell (i.e. $\mathcal{E}_{\phi}=0$,
the  $\phi^c$ field equation of motion simply reduces to $\mathcal{E}_{\phi^{c}}=0$, and in the commutative limit, we
 get $\square \phi^c=0$ as it should. Besides, the field equations (\ref{em1}) and  (\ref{em2}) are
satisfied by the vacuum solution $\phi=0$, $e^a_\mu=\partial_\mu \phi^a=\delta^a_\mu$
corresponding to the usual Moyal product. The field $\phi$ acts as a source for the noncommutativity field $\phi^c$.

The current
$\mathcal{J}^{\sigma}$ is given by
\begin{eqnarray}
\mathcal{J}^{\sigma}=\mathcal{J}^{\sigma}(0)+\mathcal{J}^{\sigma}(m^2)+\mathcal{J}^{\sigma}(\lambda)+\mathcal{J}^{\sigma}(\Omega^2)
\end{eqnarray}
where the contributions engendered by the velocity field, the mass term, the $\phi^{\star 4}$ interaction and the GW
harmonic interaction source are, respectively, expressed as
\newpage
\begin{eqnarray}
\mathcal{J}^{\sigma}(0)&=&\frac{1}{2}e\delta\phi^{a}\{\partial^{\sigma}\phi_{a},e^{-1}\}_{\star}\cr
&&+ee_{b}^{\sigma}\Big\{\frac{1}{2}\Big[ -T(\Delta)\Big(\delta\phi^{c}X_{c}\partial_{\mu}\phi_{a},\widetilde{X}^{b}\{\partial^{\mu}\phi^{a},e^{-1}\}_{\star}\Big)\nonumber\\
&&-2S(\Delta)\Big(\partial^{\mu}\phi_{a},\widetilde{X}^{b}((\delta\phi^{c}X_{c}\partial_{\mu}\phi^{a})\star e^{-1})\Big)\cr &&+2S(\Delta)\Big(\partial_{\mu}\phi_{a},\widetilde{X}^{b}(\partial^{\mu}\delta\phi^{a}\star e^{-1})\Big)\nonumber\\
&&+2T(\Delta)\Big(\delta\partial_{\mu}\phi_{a},\frac{\widetilde{X}^{b}}{2}\{\partial^{\mu}\phi^{a},e^{-1}\}_{\star}\Big)\Big]\cr
&&+\frac{1}{2}\Big[-T(\Delta)\Big(\delta\phi^{c}X_{c}\partial_{\mu}\phi,\widetilde{X}^{b}\{\partial^{\mu}\phi,e^{-1}\}_{\star}\Big)\cr
&&-2S(\Delta)\Big(\partial^{\mu}\phi,\widetilde{X}^{b}((\delta\phi^{c}X_{c}\partial_{\mu}\phi)\star
e^{-1})\Big)\Big]\cr
&&-\mathcal{L}_{\star}^{\Omega}(0)\star(\delta\phi^{b}e^{-1})+
\delta\phi^b (\mathcal{L}_\star^{\Omega}(0) \star e^{-1})\cr
 &&+T(\Delta)\Big(X_{c}(\mathcal{L}_{\star}^{\Omega}(0)),\widetilde{X}^{b}(\delta\phi^{c}e^{-1})\Big)\Big\}
\end{eqnarray}
\begin{eqnarray}
\mathcal{J}^{\sigma}(m^{2})&=&ee_{b}^{\sigma}\Big\{\frac{m^{2}}{2}\Big[
- T(\Delta)\Big(\delta\phi^a(X_a\phi), \widetilde{X}^b\{\phi ,
e^{-1}\}\Big)\cr && + 2S(\Delta)\Big(\delta\phi^a(X_a\phi)\star
e^{-1}, \widetilde{X}^b \phi\Big)\Big] \cr
&&-\mathcal{L}_{\star}^{\Omega}(m^2)\star(\delta\phi^{b}e^{-1})+
\delta\phi^b (\mathcal{L}_\star^{\Omega}(m^2) \star e^{-1})\cr
 &&+T(\Delta)\Big(X_{c}(\mathcal{L}_{\star}^{\Omega}(m^2)),\widetilde{X}^{b}(\delta\phi^{c}e^{-1})\Big)\Big\}
\end{eqnarray}
\begin{eqnarray}
\mathcal{J}^{\sigma}(\lambda)&=&ee_{b}^{\sigma}\Big\{\frac{\lambda}{4!}\Big[-T(\Delta)\Big(\delta\phi^{c}X_{c}\phi,\widetilde{X}^{b}\{\phi\star\phi,\{\phi,e^{-1}\}_{\star}\}_{\star}\Big)\cr &&-2S(\Delta)\Big(\phi,\widetilde{X}^{b}((\delta\phi^{c}X_{c}\phi)\star\phi\star\phi\star e^{-1})\Big)\nonumber\\
&&-2S(\Delta)\Big(\phi\star\phi,\widetilde{X}^{b}((\delta\phi^{c}X_{c}\phi)\star\phi\star
e^{-1})\Big)\cr &&-2S(\Delta)
\Big(\phi\star\phi\star\phi,\widetilde{X}^{b}((\delta\phi^{c}X_{c}\phi)\star
e^{-1})\Big)\Big] \cr
&&-\mathcal{L}_{\star}^{\Omega}(\lambda)\star(\delta\phi^{b}e^{-1})+
\delta\phi^b (\mathcal{L}_\star^{\Omega}(\lambda) \star e^{-1})\cr
 &&+T(\Delta)\Big(X_{c}(\mathcal{L}_{\star}^{\Omega}(\lambda)),\widetilde{X}^{b}(\delta\phi^{c}e^{-1})\Big)\Big\}
\end{eqnarray}
\begin{eqnarray}
\mathcal{J}^{\sigma}(\Omega^2)
&=&ee_{b}^{\sigma}\Big\{\frac{\Omega^{2}}{2}\Big[-T(\Delta)\Big(\delta\phi^{c}X_{c}(\tilde{x}\phi),\widetilde{X}^{b}\{\tilde{x}\phi,e^{-1}\}_{\star}\Big)\cr
&&-2S(\Delta)\Big(\tilde{x}\phi,\widetilde{X}^{b}((\delta\phi^{c}X_{c}(\tilde{x}\phi))\star
e^{-1})\Big)\Big] \cr
&&-\mathcal{L}_{\star}^{\Omega}(\Omega^{2})\star(\delta\phi^{b}e^{-1})+
\delta\phi^b (\mathcal{L}_\star^{\Omega}(\Omega^2) \star
e^{-1})\cr
 &&+T(\Delta)\Big(X_{c}(\mathcal{L}_{\star}^{\Omega}(\Omega^2)),\widetilde{X}^{b}(\delta\phi^{c}e^{-1})\Big)\Big\},
\end{eqnarray}
where
\begin{eqnarray}
&&\mathcal{L}_\star^{\Omega}(0)=\frac{1}{2}\partial_{\mu}\phi\star\partial^{\mu}\phi+\frac{1}{2}\partial_{\mu}\phi_{a}\star\partial^{\mu}\phi^{a},\quad
\mathcal{L}_\star^{\Omega}(m^2)=\frac{m^2}{2}\phi\star\phi,\quad\cr
&&\mathcal{L}_\star^{\Omega}(\lambda)=\frac{\lambda}{4!}\phi\star\phi\star\phi\star\phi,
\quad\mathcal{L}_\star^{\Omega}(\Omega^2)=\frac{\Omega^{2}}{2}(\tilde{x}_{\mu}\phi)\star(\tilde{x}^{\mu}\phi).
\end{eqnarray}

%\end{center}
 \end{enumerate}
Let us now deal with the symmetry analysis  and deduce the conserved currents.
Performing the  functional variation of the fields and  coordinate transformation
\begin{eqnarray}
 \phi'(x)=\phi(x)+\delta\phi(x),\quad
\phi'^{c}(x)=\phi^{c}(x)+\delta\phi^{c}(x),\quad
x'^{\mu}=x^{\mu}+\epsilon^{\mu}
\end{eqnarray}
and using
 $\mbox{d}^{D}x'=[1+\partial_{\mu}\epsilon^{\mu}+\textbf{O}(\epsilon^{2})]\mbox{d}^{D}x$
lead to the following  variation of the action, to first order in
 $\delta\phi(x), \delta\phi^{c}(x), \tilde{x}$ and $\epsilon^{\mu}$:
\begin{eqnarray}
\delta\mathcal{S}_{\star}^{\Omega}&=& \int\,e\mbox{d}^{D}x\Big\{\Big{\vert}\frac{\partial x'}{\partial x}
\Big{\vert}\star(\mathcal{L'}_{\star}^{\Omega}\star e^{-1})\Big\}-\int\,e\mbox{d}^{D}x\,
 (\mathcal{L}_{\star}^{\Omega}\star e^{-1})\nonumber\\
&=&\int\,\mbox{d}^{D}x\Big\{\delta\Big((\mathcal{L}_{\star}^{\Omega}\star e^{-1})e\Big)
+\partial_{\mu}\epsilon^{\mu}\star \Big((\mathcal{L}_{\star}^{\Omega}\star e^{-1})e\Big)\Big\}\nonumber\\
&=&\int{\mbox{d}}^D x\hspace{1mm}\Big\{\delta_\phi\Big((\mathcal{L}_{\star}^{\Omega}\star e^{-1})e\Big)
+\delta_{\phi^{c}}\Big((\mathcal{L}_{\star}^{\Omega}\star e^{-1})e\Big)
\cr
&&+\delta_{\tilde{x}}\Big(\mathcal{L}_{\star}^{\Omega}\star e^{-1})e\Big)+\epsilon^{\mu}\star\partial_{\mu}[(\mathcal{L}_{\star}^{\Omega}\star e^{-1})e]
\cr 
&&+\partial_{\mu}\epsilon^{\mu}\star(\mathcal{L}_{\star}^{\Omega}\star e^{-1})e\Big\}.
\end{eqnarray}
On shell, and integrated on a manifold $M$ (so that the total derivative terms do not disappear),
 we get:
\begin{eqnarray}\label{noether}
\delta\mathcal{S}_{\star}^{\Omega}= \int_{M}{\mbox{d}}^D x\hspace{1mm}
\partial_{\sigma}\Big[\mathcal{K}^{\sigma}+\mathcal{J}^{\sigma}+\mathcal{R}^{\sigma}+
\epsilon^{\sigma}\star\Big((\mathcal{L}_{\star}^{\Omega}\star e^{-1})e\Big)\Big]
\end{eqnarray}
where $\mathcal{R}^{\sigma}$
is defined as follows:
\begin{eqnarray}
\mathcal{R}^{\sigma}&=&\frac{\Omega^{2}}{8}ee_{b}^{\sigma}\Big\{T(\Delta)\Big(\delta\tilde{x},
\widetilde{X}^{b}\{\phi,\{e^{-1},\{\tilde{x},\phi\}_{\star}\}_{\star}\}_{\star}\Big)
\nonumber\\
&&+2S(\Delta)\Big(\{\tilde{x},\phi\}_{\star},\widetilde{X}^{b}(\delta\tilde{x}\star\phi\star e^{-1})\Big)\nonumber\\
&&+2S(\Delta)\Big(\{\phi,\tilde{x}\star\phi\}_{\star},\widetilde{X}^{b}(\delta\tilde{x}\star e^{-1})\Big)
\nonumber\\
&&+2S(\Delta)\Big(\phi,\widetilde{X}^{b}(\delta\tilde{x}\star\{\tilde{x},\phi\}_{\star}\star e^{-1})\Big)\Big\}.
\end{eqnarray}
Therefore the current $\mathcal{J}^{\sigma}$
reads
%\footnote{\begin{center}\begin{eqnarray}
%\mathcal{J}^{\sigma}&=&\mathcal{K}^{\sigma}(\delta\phi\rightarrow-\delta\phi^{c}X_{c}\phi)+\mathcal{R}^{\sigma}(\delta\tilde{x}\rightarrow-\delta\phi^{c}X_{c}\tilde{x})+\frac{e\delta\phi^{c}}{2}X_{c}\phi.\{\partial^{\sigma}\phi,e^{-1}\}_{\star} +\frac{e\delta\phi^{c}}{2}.\{\partial^{\sigma}\phi_{c},e^{-1}\}_{\star}\nonumber\\
%&&+ee_{b}^{\sigma}\Big\{-\mathcal{L}_{\star}^{\Omega}\star(\delta\phi^{b}e^{-1})+ \delta\phi^b (\mathcal{L}_{\star}^{\Omega} \star e^{-1})+T(\Delta)\Big(X_{c}(\mathcal{L}_{\star}^{\Omega}),\widetilde{X}^{b}(\delta\phi^{c}e^{-1})\Big)\nonumber\\
%&&+T(\Delta)\Big(\partial_{\mu}(\delta\phi^{c}e_{c}^{\rho})\partial_{\rho}\phi,\frac{1}{2}\widetilde{X}^{b}\{\partial^{\mu}\phi,e^{-1}\}_{\star}\Big)+S(\Delta)\Big(\partial_{\mu}\phi,\widetilde{X}^{b}((\partial_{\mu}(\delta\phi^{c}e_{c}^{\rho})\partial_{\rho}\phi)\star e^{-1})\Big)\nonumber\\
%&&+T(\Delta)\Big(\partial_{\mu}(\delta\phi^{c}e_{c}^{\rho})\partial_{\rho}\phi^{d},\frac{1}{2}\widetilde{X}^{b}\{\partial^{\mu}\phi_{d},e^{-1}\}_{\star}\Big)+S(\Delta)\Big(\partial_{\mu}\phi^{d},\widetilde{X}^{b}((\partial_{\mu}(\delta\phi^{c}e_{c}^{\rho})\partial_{\rho}\phi_{d})\star e^{-1})\Big)\Big\}
%                                                      \end{eqnarray}\end{center}}
%\begin{center}
\begin{eqnarray}
\mathcal{J}^{\sigma}&=&\mathcal{K}^{\sigma}(\delta\phi\rightarrow-\delta\phi^{c}X_{c}\phi)
+\mathcal{R}^{\sigma}(\delta\tilde{x}\rightarrow-\delta\phi^{c}X_{c}\tilde{x})
\cr
&&+\frac{e\delta\phi^{c}}{2}X_{c}\phi.\{\partial^{\sigma}\phi,e^{-1}\}_{\star}
+\frac{e\delta\phi^{c}}{2}.\{\partial^{\sigma}\phi_{c},e^{-1}\}_{\star}\cr &&+ee_{b}^{\sigma}\Big\{-\mathcal{L}_{\star}^{\Omega}\star(\delta\phi^{b}e^{-1})+
\delta\phi^b (\mathcal{L}_{\star}^{\Omega} \star e^{-1})
\nonumber\\
&&+T(\Delta)\Big(X_{c}(\mathcal{L}_{\star}^{\Omega}),\widetilde{X}^{b}(\delta\phi^{c}e^{-1})\Big)\cr &&+\frac{1}{2}T(\Delta)\Big(\partial_{\mu}(\delta\phi^{c}e_{c}^{\rho})\partial_{\rho}\phi,
\widetilde{X}^{b}\{\partial^{\mu}\phi,e^{-1}\}_{\star}\Big)\nonumber\\
&&+S(\Delta)\Big(\partial_{\mu}\phi,
\widetilde{X}^{b}((\partial_{\mu}(\delta\phi^{c}e_{c}^{\rho})\partial_{\rho}\phi)\star
 e^{-1})\Big)\Big\}\nonumber\\
&&+\frac{1}{2}ee_{b}^{\sigma}\Big\{-T(\Delta)\Big(\delta\phi^{c}X_{c}\partial_{\mu}\phi_{a},
\widetilde{X}^{b}\{\partial^{\mu}\phi^{a},e^{-1}\}_{\star}\Big)\nonumber\\
&&-2S(\Delta)\Big(\partial^{\mu}\phi_{a},
\widetilde{X}^{b}((\delta\phi^{c}X_{c}\partial_{\mu}\phi^{a})\star e^{-1})\Big)\cr &&+2S(\Delta)\Big(\partial_{\mu}\phi_{a},\widetilde{X}^{b}(\partial^{\mu}\delta\phi^{a}
\star e^{-1})\Big)\nonumber\\
&&+T(\Delta)\Big(\partial_{\mu}\delta\phi_{a},\widetilde{X}^{b}
\{\partial^{\mu}\phi^{a},e^{-1}\}_{\star}\Big)\Big\}.
\end{eqnarray}
$\mathcal{K}^{\sigma}$ keeps the previous defined expression.
In contrary to the result in \cite{aschieri} for ordinary $\phi_\star^4$ theory,
the twisted GW action is not invariant under global translation. Now imposing the constraint
  $\frac{\delta\mathcal{S}_{\star}^{\Omega}}{\delta\tilde{x}}=0$ giving
\begin{eqnarray}
e\frac{\Omega^{2}}{8}\{\phi,\{e^{-1},\{\tilde{x},\phi\}_{\star}\}_{\star}\}_{\star}=0,
\end{eqnarray}
coupled to the transformations
\begin{eqnarray}
\delta\phi=-\epsilon^{\nu}\partial_{\nu}\phi,\quad\delta\phi^{c}=-\epsilon^{\nu}\partial_{\nu}\phi^{c},
\quad\epsilon^{\nu}=\mbox{constant}
\end{eqnarray}
that we subtitute into (\ref{noether}) and taking into account $e_{\nu}^{a}=\partial_{\nu}\phi^{a}$, we infer from
the relation
\begin{eqnarray}
0=\delta\mathcal{S}_{\star}^{\Omega}= \int_{M}{\mbox{d}}^D x\hspace{1mm}
 \epsilon^{\nu}\partial_{\mu}\mathcal{T}^{\mu}_{\nu}
\end{eqnarray}
 the EMT
\begin{eqnarray}\label{energy}
\mathcal{T}^{\mu}_{\nu}&=&-\frac{e}{2}(\partial_{\nu}\phi)\{\partial^{\mu}\phi,e^{-1}\}_{\star}
-\frac{e}{2}(\partial_{\nu}\phi_{c})\{\partial^{\mu}\phi^{c},e^{-1}\}_{\star}
\cr
&&+ee_{b}^{\mu}\Big\{\mathcal{L}_{\star}^{\Omega}\star(e^{-1}\partial_{\nu}\phi^{b})+T(\Delta)\Big(X_{c}\mathcal{L}_{\star}^{\Omega},\widetilde{X}^{b}(e^{-1}\partial_{\nu}\phi^{c})\Big)\cr
&&+\Omega^{2}\Theta^{-1}_{\gamma\nu}\Big[S(\Delta)\Big(\{\tilde{x}^{\gamma},
\phi\}_{\star},\widetilde{X}^{b}(\phi\star e^{-1})\Big)\nonumber\\
&&+S(\Delta)\Big(\{\phi,\tilde{x}^{\gamma}\star\phi\}_{\star},\widetilde{X}^{b}(e^{-1})\Big)\cr
&&+S(\Delta)\Big(\phi,\widetilde{X}^{b}\{\tilde{x}^{\gamma},\phi\}_{\star}\star e^{-1}\Big)\Big]\Big\}.
\end{eqnarray}
This tensor is neither symmetric nor locally conserved.
 In the case of standard Moyal  product, it  reduces
 to the NC EMT computed in \cite{Ben-Houk1} and its regularization can be worked out in the same way as done
 in that work.
Similary,  the transformation
\begin{eqnarray}
\delta\phi=-\epsilon^{\nu}\partial_{\nu}\phi=-\epsilon^{\nu\rho}x_{\rho}\partial_{\nu}\phi,\quad
\delta\phi^{c}=-\epsilon^{\nu}\partial_{\nu}\phi^{c}=-\epsilon^{\nu\rho}x_{\rho}\partial_{\nu}\phi^{c},\quad
\epsilon^{\nu}=\epsilon^{\nu\rho}x_{\rho}
\end{eqnarray}
with $\epsilon^{\nu\rho}$ an infinitesimal constant skew symmetric Lorentz parameter,
and  $\epsilon^{\nu\rho}x_{[\nu}\partial_{\rho]}\phi=-2\epsilon^{\nu\rho}x_{\rho}\partial_{\nu}\phi$,
substituted into (\ref{noether}) yields
\begin{eqnarray}
0=\delta\mathcal{S}_{\star}^{\Omega}= \int_{M}{\mbox{d}}^D x\hspace{1mm} \epsilon^{\nu\rho}\partial_{\mu}\mathcal{M}^{\mu}_{\nu\rho},
\end{eqnarray}
which affords the AMT as
\begin{eqnarray}
\mathcal{M}^{\mu}_{\nu\rho}&=&\frac{e}{4}x_{[\nu}\partial_{\rho]}\phi\{\partial^{\mu}\phi,e^{-1}\}_{\star}+\frac{e}{4}x_{[\nu}\partial_{\rho]}\phi_{c}\{\partial^{\mu}\phi^{c},e^{-1}\}_{\star}\cr &&-\frac{ee_{b}^{\mu}}{2}\Big(\mathcal{L}_{\star}^{\Omega}\star(e^{-1}x_{[\nu}\partial_{\rho]}\phi^{b})\Big) \nonumber\\
&&+\frac{ee_{b}^{\mu}}{2}\Big\{T(\Delta)\Big(X_{c}\mathcal{L}_{\star}^{\Omega},\widetilde{X}^{b}(e^{-1}x_{[\nu}\partial_{\rho]}\phi^{c})\Big)\cr
&&-T(\Delta)\Big(\partial_{[\nu}\phi,\frac{1}{2}\widetilde{X}^{b}(\{\partial_{\rho]}\phi,e^{-1}\}_{\star})\Big)\nonumber\\
&&-T(\Delta)\Big(\partial_{[\nu}\phi^{d},\frac{1}{2}\widetilde{X}^{b}(\{\partial_{\rho]}\phi_{d},e^{-1}\}_{\star})\Big)\cr &&+S(\Delta)\Big(\partial_{[\nu}\phi,\widetilde{X}^{b}(\partial_{\rho]}\phi\star e^{-1})\Big)\nonumber\\
&&+S(\Delta)\Big(\partial_{[\nu}\phi_{d},\widetilde{X}^{b}(\partial_{\rho]}\phi^{d}\star e^{-1})\Big) \nonumber\\
&&-\frac{\Omega^{2}}{4}\Theta^{-1}_{\gamma[\nu}\Big[T(\Delta)\Big(x_{\rho]},\widetilde{X}^{b}(\{\phi,\{e^{-1},\{\tilde{x}^{\gamma},\phi\}_{\star}\}_{\star}\}_{\star})\Big)\nonumber\\
&&+2S(\Delta)\Big(\{\tilde{x}^{\gamma},\phi\}_{\star},\widetilde{X}^{b}(x_{\rho]}\star\phi\star e^{-1})\Big)\cr &&+2S(\Delta)\Big(\{\phi,\tilde{x}^{\gamma}\star\phi\}_{\star},\widetilde{X}^{b}(x_{\rho]}\star e^{-1})\Big)\nonumber\\
&&+2S(\Delta)\Big(\phi,\widetilde{X}^{b}(x_{\rho]}\star\{\tilde{x},\phi\}_{\star}\star e^{-1})\Big)\Big]\Big\}.
\end{eqnarray}
This angular momentum tensor is not conserved, in contrary to the result obtained for the non renormalizable twisted
$\phi^{\star 4}$ model studied in \cite{aschieri}. This analysis is compatible with the previous GW model
investigation \cite{bghnk0}. One recovers the canonical angular momentum tensor of the decoupled fields
in the commutative limit.
Defining now
the dilatation transformation by
\begin{eqnarray}
x\rightarrow x'=\epsilon x;\quad\phi(x)\rightarrow \phi'(x')=\phi'(\epsilon x)=\epsilon^{-\Delta}\phi(x),
\end{eqnarray}
where $\epsilon$ is a constant number, and $\Delta$ is the scale dimension of the field $\phi$,  we note that
 the GW action is invariant over dilatation symmetry if $\Delta=0$ and $\epsilon=\pm 1$, implying
\begin{eqnarray}
 x'=x,\quad \phi'(x)=\phi(x);\quad\mbox{ or }\quad x'=-x, \quad \phi'(-x)=\phi(x)
\end{eqnarray}
which is nothing but a parity transformation  of the space-time inducing a  conserved current:
\begin{eqnarray}
\mathcal{D}^{\mu} =\mathcal{R}^{\mu}(\delta\tilde{x}\rightarrow -2\tilde{x})
-2x^{\mu}(\mathcal{L}_{\star}^{\Omega}\star e^{-1})e.
\end{eqnarray}
 Finally, the
 EMT, AMT and DC can be computed
 under the well defined field values at the boundary, i.e.  $\int\,e\mbox{d}^{D}x\,X_{b} S(\Delta)(f,\widetilde{X}^{b}g)=0$, to give
  simplified expressions. In this case, there follow
\begin{eqnarray}\label{energy}
\mathcal{T}^{\mu}_{\nu}&=&-\frac{e}{2}(\partial_{\nu}\phi)\{\partial^{\mu}\phi,e^{-1}\}_{\star}-\frac{e}{2}(\partial_{\nu}\phi_{c})\{\partial^{\mu}\phi^{c},e^{-1}\}_{\star}\cr
&&+ee_{b}^{\mu}\Big\{\mathcal{L}_{\star}^{\Omega}\star(e^{-1}\partial_{\nu}\phi^{b})+T(\Delta)\Big(X_{c}\mathcal{L}_{\star}^{\Omega},\widetilde{X}^{b}(e^{-1}\partial_{\nu}\phi^{c})\Big),
\end{eqnarray}
and
%\newpage
\begin{center}
\begin{eqnarray}
\mathcal{M}^{\mu}_{\nu\rho}&=&\frac{e}{4}x_{[\nu}\partial_{\rho]}\phi\{\partial^{\mu}\phi,e^{-1}\}_{\star} +\frac{e}{4}x_{[\nu}\partial_{\rho]}\phi_{c}\{\partial^{\mu}\phi^{c},e^{-1}\}_{\star}\cr &&-\frac{ee_{b}^{\mu}}{2}\Big(\mathcal{L}_{\star}^{\Omega}\star(e^{-1}x_{[\nu}\partial_{\rho]}\phi^{b})\Big)\cr &&+\frac{ee_{b}^{\mu}}{2}\Big\{T(\Delta)\Big(X_{c}\mathcal{L}_{\star}^{\Omega},\widetilde{X}^{b}(e^{-1}x_{[\nu}\partial_{\rho]}\phi^{c})\Big)\cr &&-T(\Delta)\Big(\partial_{[\nu}\phi,\frac{1}{2}\widetilde{X}^{b}(\{\partial_{\rho]}\phi,e^{-1}\}_{\star})\Big)\cr &&-T(\Delta)\Big(\partial_{[\nu}\phi^{d},\frac{1}{2}\widetilde{X}^{b}(\{\partial_{\rho]}\phi_{d},e^{-1}\}_{\star})\Big)\cr  &&-\frac{\Omega^{2}}{4}\Theta^{-1}_{\gamma[\nu}T(\Delta)\Big(x_{\rho]},\widetilde{X}^{b}(\{\phi,\{e^{-1},\{\tilde{x}^{\gamma},\phi\}_{\star}\}_{\star}\}_{\star})\Big)\Big\}
\end{eqnarray}
\end{center}
and the current of dilatation symmetry in the form
\begin{eqnarray}
\mathcal{D}^{\mu}&=&-\Omega^{2}ee_{b}^{\mu}T(\Delta)\Big(\tilde{x},\widetilde{X}^{b}\{\phi,
\{e^{-1},\{\tilde{x},\phi\}_{\star}\}_{\star}\}_{\star})\Big)\cr 
&&-2x^{\mu}(\mathcal{L}_{\star}^{\Omega}\star e^{-1})e.
\end{eqnarray}
\section{Concluding remarks}
The following features are worthy of attention from this study:
\begin{enumerate}
 \item  The ordinary $\phi^{4}-$theory leads to nonlocally conserved and symmetric EMT
 and AMT \cite{Gerhold} while the twisted non renormalizable $\phi^{4}-$theory \cite{aschieri} restores the local conservation of these tensors because of nonzero boundary conditions.

\item Both ordinary GW \cite{Ben-Houk1,bghnk0} and twisted GW models provide nonlocally conserved and nonsymmetric EMT, AMT and DC due to the presence of the harmonic term $\Omega$.
\end{enumerate}
As shown in \cite{Ben-Houk1}, all these physical quantities can be subjected to well known Jackiw and Wilson regularization
 procedures to acquire the local conservation property.

 \section*{Acknowledgements}
This work is partially supported by the ICTP through the OEA-ICMPA-Prj-15.
The ICMPA is in partnership with the Daniel Iagolnitzer foundation (DIF), France.
The authors thank Dr J. Ben Geloun and Marija Dimitrijevic for fruitful discussions.
%\end{large}

\section*{Appendix}
We summarize here useful properties  of the dynamical $\star$-product
 expanded as
\begin{eqnarray}\label{starpro}
f\star g&=&fg+\frac{i}{2}\Theta^{ab}X_{a}fX_{b}g\cr &&+\frac{1}{2!}\Big(\frac{i}{2}\Big)^{2}\Theta^{a_{1}b_{1}}\Theta^{a_{2}b_{2}}(X_{a_{1}}X_{a_{2}}f)(X_{b_{1}}X_{b_{2}}g) +\cdots\nonumber\\
&\equiv & e^{\Delta}(f,g)
\end{eqnarray}
where powers of the bilinear operator $\Delta$ are defined as
\begin{eqnarray}
\Delta(f,g)=\frac{i}{2}\Theta^{ab}(X_{a}f)(X_{b}g),\quad \Delta^{0}(f,g)=fg\nonumber\\
\Delta^{n}(f,g)=\Big(\frac{i}{2}\Big)^{n}\Theta^{a_{1}b_{1}}\cdots\Theta^{a_{n}b_{n}}(X_{a_{1}}
\cdots X_{a_{n}}f)(X_{b_{1}}\cdots X_{b_{n}}g).
\end{eqnarray}
From the definition (\ref{starpro}) we deduce the following identities
 (straightforward generalization of the  usual Moyal product identities):
\begin{eqnarray}
&f\star g=fg+X_{a}T(\Delta)(f,\widetilde{X}^{a}g)\\
&[f,g]_{\star}=f\star g-g\star f =2X_{a}S(\Delta)(f,\widetilde{X}^{a}g)\\
&\{f,g\}_{\star}=f\star g+ g\star f=2fg+2X_{a}R(\Delta)(f,\widetilde{X}^{a}g)
\end{eqnarray}
where
\begin{eqnarray}
T(\Delta)&=&\frac{e^{\Delta}-1}{\Delta},\quad S(\Delta)=\frac{sinh(\Delta)}{\Delta}, \nonumber
\\ R(\Delta)&=&\frac{cosh(\Delta)-1}{\Delta}\mbox{ and } \widetilde{X}^{a}=\frac{i}{2}\Theta^{ab}X_{b}
\end{eqnarray}
implying that $S(\Delta)(.,\widetilde{X}.)$ is a bilinear antisymmetric operator and
\begin{eqnarray}
 T(\Delta)(f,\widetilde{X}^{a}g)-T(\Delta)(g,\widetilde{X}^{a}f)=2S(\Delta)(f,\widetilde{X}^{a}g).
\end{eqnarray}
The formulas of derivatives and variations are given  by \cite{aschieri}
\begin{eqnarray}
\delta_{\phi^{c}}e_{a}^{\mu}=-e_{b}^{\mu}X_{a}(\delta\phi^{b}),\quad\partial_{\mu}e=eX_{a}(\partial_{\mu}\phi^{a}),\cr
\delta_{\phi^{c}}e=eX_{a}(\delta\phi^{a}),\quad
 \delta_{\phi^{c}}e^{-1}=-e^{-1}X_{a}\delta(\phi^{a}),\cr
\delta_{\phi^{c}}X_{a}=-X_{a}(\delta\phi^{b})X_{b},\quad
eX_{a}(f)=\partial_{\mu}(ee_{a}^{\mu}f)
\end{eqnarray}
To compute $\delta_{\phi^{c}}$ variations, the following identity is useful:
\begin{eqnarray}\label{111}
 \delta_{\phi^{c}}(f\star g)=-(\delta\phi^{c}X_{c}f)\star
  g-f\star( \delta\phi^{c}X_{c}g)+\delta\phi^{c}X_{c}(f\star g),
\end{eqnarray}
where the functions $f$ and $g$ do not depend on $\phi^{c}$.
 By induction, one can immediately prove  that (\ref{111}) holds for $\star$-products of an arbitrary number of factors:
\begin{eqnarray}
\delta_{\phi^{c}}(f\star g\star \cdots \star h)&=&-(\delta\phi^{c}X_{c}f)\star g\star\cdots\star h\cr &&-f\star(\delta\phi^{c}X_{c}g)\star\cdots\star h \nonumber\\
&&-\cdots -f\star g\star\cdots\star(\delta\phi^{c}X_{c}h)\cr
&&+\delta\phi^{c}X_{c}(f\star g\star\cdots\star h)
\end{eqnarray}
%Note that the cyclicity under the integral sign is prohibited, (even with suitable boundary conditions at infinity),
% i.e.
One has also:
\begin{eqnarray}
\int\,\mbox{d}^{D}x \, (f\star g)\neq \int\,\mbox{d}^{D}x \, (g\star f),
\end{eqnarray}
but
\begin{eqnarray}
 \int\,e\mbox{d}^{D}x \, (f\star g)=\int\,e\mbox{d}^{D}x(fg)=\int\,e\mbox{d}^{D}x \, (g\star f).
\end{eqnarray}
%so that up to boundary terms:
%\begin{eqnarray}
%\int\,e\mbox{d}^{D}x \, (f\star g)=\int\,e\mbox{d}^{D}x(fg)=\int\,e\mbox{d}^{D}x \, (g\star f)
%\end{eqnarray}

\end{titlepage}
\end{document}